\documentclass[iop]{emulateapj}
\usepackage{apjfonts}
\usepackage{rotating}
\usepackage{hyperref}

\slugcomment{Accepted for publication in \apj}

\shorttitle{Stellar mass structure to $z=2.5$}
\shortauthors{Szomoru et al.}

\begin{document}

\title{The stellar mass structure of massive galaxies from $z=0$ to $z=2.5$; surface density profiles and half-mass radii}

\author{
Daniel Szomoru\altaffilmark{1},
Marijn Franx\altaffilmark{1},
Pieter~G.~van Dokkum\altaffilmark{2},
Michele Trenti\altaffilmark{3}$^,$\altaffilmark{4},
Garth~D.~Illingworth\altaffilmark{5},
Ivo Labb\'e\altaffilmark{1},
Pascal Oesch\altaffilmark{5}$^,$\altaffilmark{6},
}

\altaffiltext{1}
{Leiden Observatory, Leiden University, P.O. Box 9513, 2300 RA Leiden, The Netherlands}
\altaffiltext{2}
{Department of Astronomy, Yale University, New Haven, CT 06520-8101, USA}
\altaffiltext{3}
{Institute of Astronomy, University of Cambridge, Madingley Road, Cambridge, CB3 0HA, United Kingdom}
\altaffiltext{4}
{Kavli Fellow}
\altaffiltext{5}
{UCO/Lick Observatory, University of California, Santa Cruz, CA 95064, USA}
\altaffiltext{6}
{Hubble Fellow}

\begin{abstract}
We present stellar mass surface density profiles of a mass-selected sample of 177 galaxies at $0.5 < z < 2.5$, obtained using very deep HST optical and near-infrared data over the GOODS-South field, including recent CANDELS data. Accurate stellar mass surface density profiles have been measured for the first time for a complete sample of high-redshift galaxies more massive than $10^{10.7} M_\odot$.  The key advantage of this study compared to previous work is that the surface brightness profiles are deconvolved for PSF smoothing, allowing accurate measurements of the structure of the galaxies. The surface brightness profiles account for contributions from complex galaxy structures such as rings and faint outer disks. Mass profiles are derived using radial rest-frame $u−g$ color profiles and a well-established empirical relation between these colors and the stellar mass-to-light ratio. We derive stellar half-mass radii from the mass profiles, and find that these are on average $\sim25\%$ smaller than rest-frame $g$ band half-light radii. This average size difference of 25\% is the same at all redshifts, and does not correlate with stellar mass, specific star formation rate, effective surface density, S\'ersic index, or galaxy size.  Although on average the difference between half-mass size and half-light size is modest, for approximately 10\% of massive galaxies this difference is more than a factor two. These extreme galaxies are mostly extended, disk-like systems with large central bulges. These results are robust, but could be impacted if the central dust extinction becomes high. ALMA observations can be used to explore this possibility. These results provide added support for galaxy growth scenarios wherein massive galaxies at these epochs grow by accretion onto their outer regions.
\end{abstract}


\section{Introduction}

Over the past decades quantitative studies of high-redshift galaxy structure have advanced tremendously. Sensitive, high resolution instruments such as the Hubble Space Telescope's (\textit{HST}) Advanced Camera for Surveys (ACS) and the Wide Field Camera 3 (WFC3) have made it possible to measure the structure of faint high-redshift galaxies at sub-kpc scales. Furthermore, the availability of easy-to-use photometric redshift and stellar population fitting packages has made it possible to straightforwardly measure a wide variety of parameters for ever-increasing numbers of galaxies.

Since a small amount of recent star formation can have a disproportionately large contribution to a galaxy's light compared to its mass, galaxies are usually observed at the redmost wavelengths, where emission from young stars is weakest. At low redshift, this can be done quite effectively, since rest-frame near-infrared (NIR) data is available at high enough resolution. At higher redshifts, however, it is impossible to observe at such long wavelengths with sufficiently high angular resolution. Until recently, the Hubble Space Telescope (\textit{HST}) Advanced Camera for Surveys (ACS) was the only wide-field instrument capable of measuring the structure of high-redshift galaxies in any detail. At $z=2$, the reddest filter available on ACS, $Z_{850}$, corresponds to rest-frame near-ultraviolet (NUV) wavelengths. Use of such short-wavelength data has been shown to result in drastically different conclusions about galaxy structure and morphology, compared to rest-frame optical data (e.g., \citealt{lab03}; \citealt{tof05}; \citealt{cam11}).

With the introduction of the \textit{HST} Wide Field Camera 3 (WFC3) it has become possible to measure rest-frame optical light of $z\sim2$ galaxies at a resolution approaching that of the ACS. The redder light detected by this instrument provides a much better proxy for stellar mass. However, color gradients are known to exist to some extent in all types of galaxies at redshifts up to at least $z\sim 3$, such that most galaxies contain a relatively red core and blue outer regions (e.g., \citealt{dok10}; \citealt{szo11}; \citealt{guo11}). These color variations are caused by a combination of varying dust content, metallicity and stellar age, and imply that the stellar mass-to-light ratio ($M_*/L$) of a given galaxy varies with position within that galaxy. Thus, even though rest-frame optical light is a better tracer of stellar mass than rest-frame NUV light, neither accounts for the complexity of stellar population variations within galaxies.

By fitting stellar population models to resolved galaxy photometry, it is in principle possible to infer spatial variations in stellar mass, age, metallicity, dust content, and other parameters. This approach is currently somewhat limited by the lack of high-resolution data at infrared wavelengths, but can nonetheless be used to measure several basic stellar population properties. An example of this technique is presented in \cite{wuy12}, who have performed stellar population modeling on resolved \textit{HST} data, using integrated IR observations as constraints on the overall properties of their galaxies. In this approach the integrated photometry serves as an important tool to constrain the overall spectral energy distribution (SED) of a galaxy, while the \textit{HST} data provide information regarding the spatial variation of the stellar populations within these constraints.

In this Paper we explore an alternative method to recover $M_*/L$ variations, using a simple empirical relation between rest-frame $u-g$ color and $M_*/L$. Using this method we construct stellar mass surface density profiles corrected for the effects of the PSF, for a mass-selected sample of galaxies between $z=0$ and $z=2.5$. We compare the resulting half-mass radii to half-light radii based on rest-frame optical imaging. All sizes presented in this Paper are circularized sizes: $r_e = r_{e,a}\sqrt{b/a}$. Throughout the Paper we assume a $\Lambda$CDM cosmology with $\Omega_m = 0.3$, $\Omega_\Lambda = 0.7$, and $H_0 = 70$ km s$^{-1}$ Mpc$^{-1}$.

\section{Data and sample selection}\label{sec:data}

\subsection{HST imaging}

We make use of deep near-IR imaging of the GOODS-South field, obtained with \textit{HST}/WFC3 as part of the Cosmic Assembly Near-infrared Deep Extragalactic Legacy Survey (CANDELS; \citealt{gro11}, \citealt{koe11}). When completed, this survey will cover $\sim700$ arcmin$^2$ to 2 orbit depth in $I_{814}$, $J_{125}$ and $H_{160}$ (COSMOS, EGS and UDS fields), as well as $\sim120$ arcmin$^2$ to 12 orbit depth (GOODS-South and GOODS-North fields). We use the deepest currently available data, which consist of 9 orbits in $J_{125}$ and $H_{160}$ taken over GOODS-South. These NIR data are combined with deep \textit{HST}/ACS data in the $B_{435}$, $V_{606}$, $I_{775}$ and $Z_{850}$ filters from the Great Observatories Origins Deep Survey (GOODS ACS v2.0; \citealt{gia04}). The full width at half-maximum (FWHM) of the point-spread function (PSF) is $\approx0.12 - 0.18$ arcsec for the WFC3 observations and $\approx0.1$ arcsec for the ACS observations. The WFC3 and ACS images have been drizzled to pixel scales of 0.06 and 0.03 arcsec pixel$^{-1}$, respectively (see \citealt{koe11} for a detailed description of the CANDELS data reduction, and \citealt{gia04} for details of the ACS reduction).

We select galaxies using the $K_s$-selected FIREWORKS catalog \citep{wuy08}. This catalog combines observations of the Chandra Deep Field South (CDFS) ranging from ground-based $U$-band data to \textit{Spitzer} 24$\mu$m data. It includes spectroscopic redshifts where available, as well as photometric redshifts derived using EAZY \citep{bra08}. The photometric redshifts have a median $\Delta z/(1+z) = -0.001$ with a normalized median absolute deviation of $\sigma_{NMAD} = 0.032$ \citep{wuy08}. Stellar masses are estimated from SED fits to the full photometric data set (N.M. F\"orster Schreiber et al. 2012, in preparation), assuming a Kroupa IMF and the stellar population models of \cite{bru03}. Star formation rates have been calculated using the UV and 24 $\mu$m fluxes \citep{wuy09}.

We limit our analysis to galaxies with redshifts between $z=0.5$ and $z=2.5$; within this wavelength range we have the wavelength coverage needed to measure rest-frame $u$ and $g$ band photometry. We select galaxies with stellar masses above $10^{10.7} M_\odot$, which is the completeness limit in this redshift range \citep{wuy09}. This redshift and mass cut results in a sample of 177 galaxies, of which 110 are at $0.5 < z < 1.5$ and 67 at $1.5 < z < 2.5$.

\subsection{SDSS imaging}

We compare our high-redshift galaxies to low-redshift galaxies observed as part of the Sloan Digital Sky Survey (SDSS; \citealt{aba09}). In order to obtain the deepest possible galaxy photometry we limit our analysis to data from SDSS stripe 82 \citep{ann11}. This region of the sky has been repeatedly imaged by SDSS, resulting in data which are $\sim2$ magnitudes deeper compared to standard SDSS imaging. The PSF FWHM is $\approx0.6$ arcsec, and the images have a pixel scale of 0.396 arcsec pixel$^{-1}$.

We base our galaxy selection on the NYU Value-Added Galaxy Catalog (\citealt{bla05}); stellar masses and star formation rates are taken from the MPA-JHU catalogs\footnote{See \url{http://www.mpa-garching.mpg.de/SDSS/DR7/mass_comp.html} for a comparison between these masses and masses based on spectral indices} \citep{bri04}. We select galaxies with spectroscopically measured redshifts in a narrow redshift range $z=0.06\pm0.005$ and with stellar masses above $10^{10.7} M_\odot$. This results in a sample of 220 galaxies.

\section{Analysis}\label{sec:analysis}

\subsection{Rest-frame surface brightness profiles}

Most studies of galaxy structure at high redshift are based on parametrized two-dimensional surface brightness profile fits (e.g., using the GALFIT package of \citealt{pen02}). However, such methods do not to account for deviations from the assumed model profile, which is generally an $r^{1/n}$ S\'ersic profile. The technique used in this paper, which is described in more detail in \cite{szo10} and \cite{szo12}, is different in the sense that these deviations are explicitly included in the measurement process. The intrinsic profile is derived by first fitting a S\'ersic model profile convolved with the PSF to the observed flux, using a PSF constructed from unsaturated bright stars in the image. The residuals from this fit are then measured in radially concentric ellipses which follow the geometry of the best-fit S\'ersic profile. By adding these residuals to the best-fit S\'ersic profile, we effectively perform a first-order correction for deviations from the model profile and are able to account for complex substructures such as rings and faint outer disks. The resulting profiles closely follow the true intrinsic galaxy profiles, as shown in \cite{szo10} and \cite{szo12}.

The surface brightness profiles of all galaxies in the high-redshift sample are measured in the $B_{435}$, $V_{606}$, $I_{775}$, $Z_{850}$, $J_{125}$ and $H_{160}$ filter; this ensures sufficient wavelength coverage to accurately measure rest-frame $u-g$ colors. The SDSS galaxy profiles are measured in the $u$, $g$, $r$ and $z$ bands. Errors in the flux profiles are calculated by adding the formal flux errors, sky variance, and the estimated error in the sky background determination. The radial extent of the profiles is mostly limited by uncertainties in the sky background estimation; typically, the profiles are accurate out to radii of approximately 10 kpc. At larger radii the profiles are extrapolated using the S\'ersic model profile. The extrapolated part of the profile typically contains $\sim5-10\%$ of the total flux, depending on the galaxy profile shape.

Rest-frame $u$ and $g$ band profiles are derived by interpolating between the observed fluxes at each radius, using the SED-based interpolation package InterRest \citep{tay09}. This package uses a set of template SEDs to interpolate between observed fluxes. The resulting rest-frame fluxes thus take into account the shape of a galaxy's SED and the filter throughputs of both the observed and rest-frame filters.

\subsection{From colors to mass-to-light ratios}\label{subsec:colorml}

\begin{figure*}
\epsscale{1.2}
\plotone{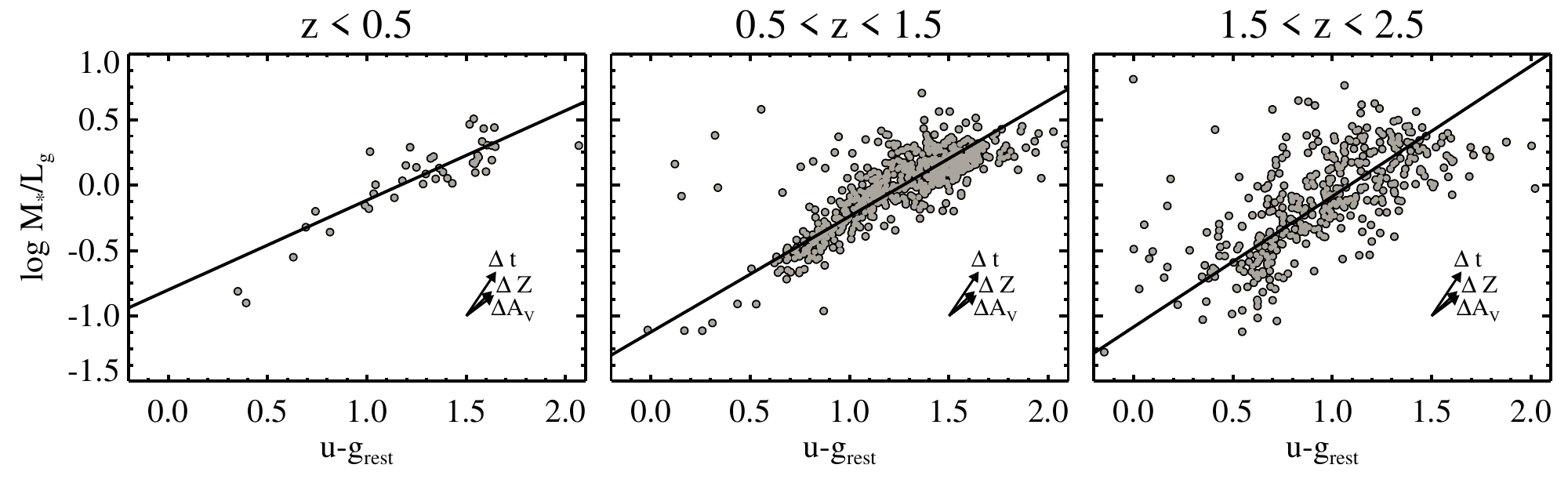}
\caption{The relationship between rest-frame $u-g$ color and stellar mass-to-light ratio in three redshift bins. Grey circles denote integrated colors and mass-to-light ratios of individual galaxies in the Chandra Deep Field South, based on data from the FIREWORKS catalog \citep{wuy08}. The best-fit linear relations are indicated by black lines. The slope of the relation is close to unity at high redshift, and flattens off at lower redshifts. The effects of age, dust extinction and metallicity on $u-g$ and $M_*/L_g$ are indicated by the arrows labeled $\Delta t$, $\Delta A_V$ and $\Delta Z$, respectively. The length of the vectors indicate the shift caused by an increase in stellar age from 1 Gyr to 5 Gyr, $1 A_V$ of dust extinction, and an increase in metallicity from 0.02 to 2.5 times solar metallicity. Increases in stellar age, dust content, and metallicity all result in a shift roughly parallel to the best-fit relation, and are thus implicitly included in our empirical relation.}\label{fig:mlcolor}
\end{figure*}

Stellar mass-to-light ratios are estimated using an empirical relation between $u_{336}-g_{475}$ and $M_*/L_g$. The $u$ and $g$ filters straddle the Balmer and 4000\AA ~breaks; $u-g$ color is therefore sensitive to changes in stellar age, dust, and metallicity. Our empirical $u-g$ --- $M_*/L_g$ relation is based on FIREWORKS photometry of galaxies in the CDFS, shown in Figure~\ref{fig:mlcolor} \citep{wuy08}. In this Figure we plot the integrated rest-frame $u-g$ colors and $\log{M_*/L_g}$ of galaxies in three redshift bins (grey circles). Rest-frame colors have been calculated using InterRest \citep{tay09}, and the mass-to-light ratios are obtained from the SED fits described in Section~\ref{sec:data}. The black lines indicate the best-fit linear relation at each redshift. The slope of the relation is 1.0 at $z\sim2$, 0.9 at $z\sim1$, and 0.7 at $z\sim0$. The uncertainty in $\log{M_*/L_g}$, given a value for $u-g$, is approximately 0.28 at $z\sim2$, and 0.13 at lower redshifts.

The reason that these relations exist is due to the fact that stellar population variations (i.e., changes in stellar age, metallicity, and dust content) all produce roughly the same effects in the $u-g$ --- $M_*/L_g$ plane (\cite{bel01}). This is indicated by the arrows in each panel of Figure~\ref{fig:mlcolor}, which all point roughly in the same direction. Due to this degeneracy we can not distinguish between dust, age, and metallicity effects, but $M_*/L_g$ can be determined quite robustly. It should also be noted that the axes in Figure~\ref{fig:mlcolor} are not completely independent, since the rest-frame $u-g$ color enters into the stellar population fit from which $M_*/L_g$ is determined.

There is significant scatter around the empirical relation used to convert $u-g$ color to $M_*/L$. Although age, metallicity, and dust variations produce similar shifts in the color-$M/L$ plane, their effects are not exactly parallel to our empirical relation. This could lead to a systematic underestimate of the mass-to-light ratios in galaxy regions that are relatively metal-rich or old, and an overestimate in regions that are relatively metal-poor or young. Since the central regions of galaxies generally contain older stars, the inferred $M_*/L$ gradients would then be too shallow. This effect is likely small, since the vectors shown in Figure~\ref{fig:mlcolor} do not diverge very strongly, and the magnitude of the shift is small for moderate stellar population differences.

\subsection{Mass profile derivation}

\begin{figure*}
\epsscale{1.2}
\plotone{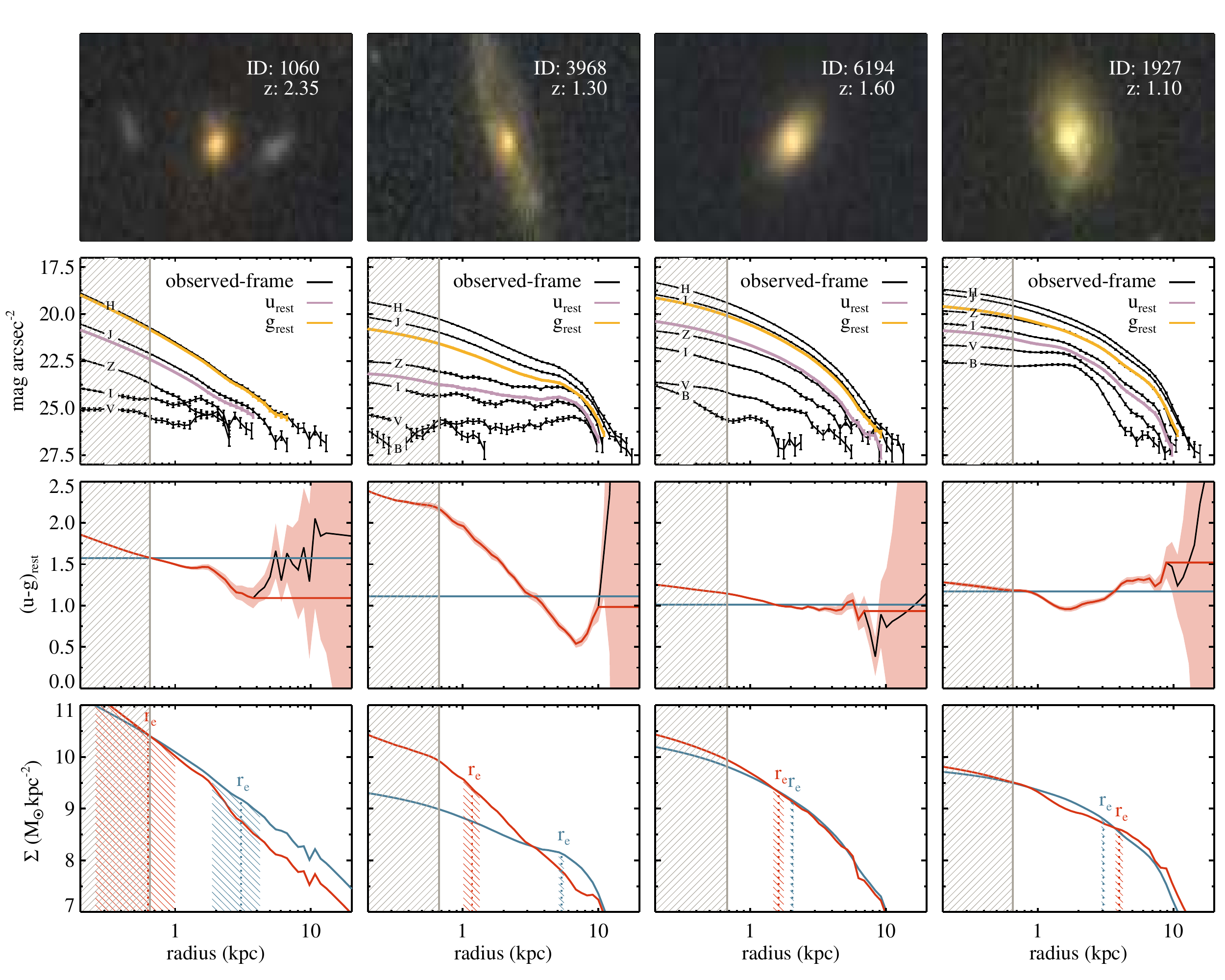}
\caption{Illustration of the conversion of observed surface brightness profiles to stellar mass surface density profiles, for four sample galaxies at $0.5 < z < 2.5$. Top row: color images, composed of rest-frame $u_{336}$, $B_{438}$ and $g_{475}$ images. Second row: observed-frame residual-corrected surface brightness profiles (shown in greyscale), with rest-frame $u$ and $g$ band profiles overlaid in purple and yellow, respectively. Third row: measured $u-g$ color profiles and average $u-g$ colors (in red and blue, respectively). The grey hatched area indicates the PSF HWHM. Bottom row: resulting surface density profiles, with color coding corresponding to the color profiles in the third row. Effective radii are shown for both profiles; the red and blue hatched areas indicate the $1-\sigma$ errors. Observed- and rest-frame photometry is generally of very high quality, and provides accurate $u-g$ color profiles out to $\sim10$ kpc. The colors at the largest radii, where the flux profiles are not reliable, are extrapolated from the last well-measured color. Half-mass radii are in some cases more than 50\% smaller than half-light radii.}\label{fig:example}
\end{figure*}

The process of deriving stellar mass surface density profiles is illustrated for four galaxies in Figure~\ref{fig:example}. The galaxies are selected to show a range of color gradients. For each galaxy we show, from top to bottom, a rest-frame $ubg$ color image, the observed-frame and rest-frame surface brightness profiles, the rest-frame $u-g$ color profile, and the resulting stellar mass surface density profile. The extent of the PSF half width at half-maximum (HWHM) is indicated by grey hatched areas.

The observed-frame residual-corrected surface brightness profiles, shown in grey-scale in the second row, are generally of high quality. The profiles deteriorate somewhat in the bluer bands for the highest-redshift galaxies. However, the profiles that are used for interpolating to rest-frame $u$ and $g$ wavelengths (i.e., measured in the bands directly red- and blueward of the rest-frame $u$ and $g$ wavelengths) have high signal-to-noise. The observed surface brightness profiles, and by extension the rest-frame $u-g$ profiles, are generally accurate out to $\sim10$ kpc.

In the third row of Figure~\ref{fig:example} we plot the observed $u-g$ color profiles as well as the average $u-g$ colors for the entire galaxy (in red and blue, respectively). The color gradients range from very steep ($\Delta(u-g)/\Delta{\log{r}} = -1$) to very shallow. At large radii ($r \gtrsim 10$ kpc) the color profiles become extremely uncertain. Due to the low surface brightness at these radii, the ratio of $u$ band to $g$ band flux is very sensitive to small errors in either flux profile. This can result in colors that become unrealistically blue or red. We therefore limit the color profiles to radii where the error in $u-g$ is smaller than 0.2 dex; the $u-g$ color at larger radii is fixed to the value at the threshold radius. This should not have a strong effect on the resulting stellar mass surface density profile, since the surface densities at these large radii are so low that they contribute very little to the total mass, even for high $M_*/L_g$.

\begin{figure*}
\epsscale{1.2}
\plotone{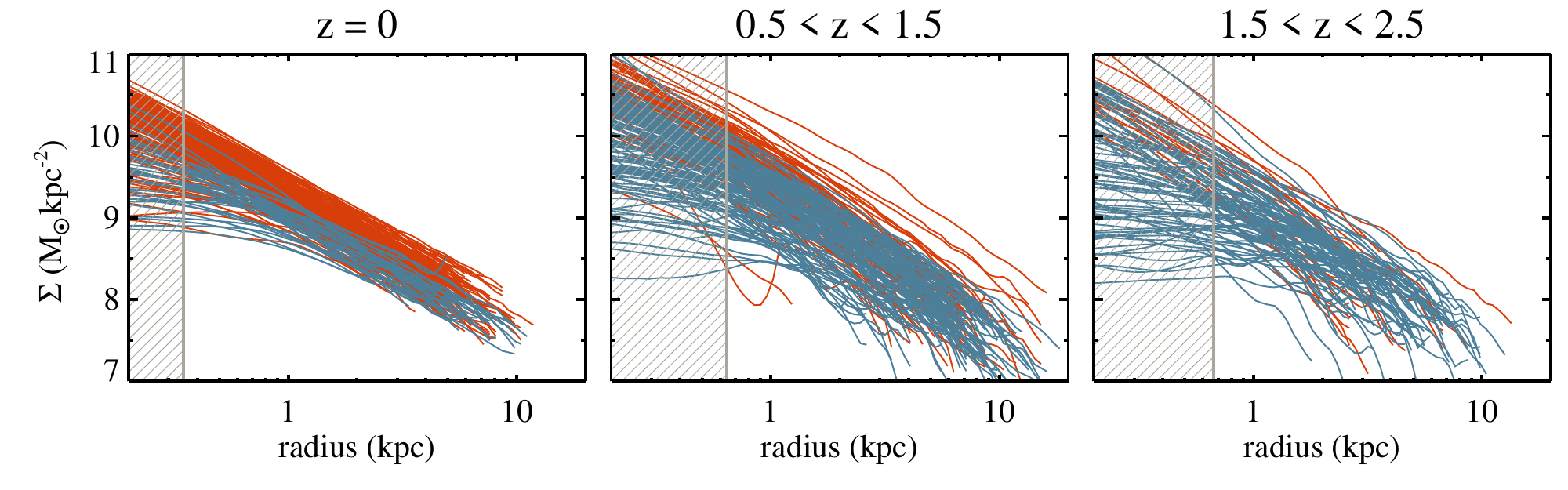}
\caption{Stellar mass surface density profiles of galaxies with $M_* > 10^{10.7} M_\odot$ at $z=0$, $z\sim1$, and $z\sim2$. Individual profiles are shown in blue and red for starforming and quiescent galaxies, respectively. The approximate PSF HWHM in each redshift bin is indicated by the hatched region. The profiles are plotted up to the radius where the errors become significant; they are accurate out to $\sim10$ kpc, and down to surface densities of around $10^{7.5} M_\odot$ kpc$^{-2}$.}\label{fig:profiles}
\end{figure*}

The resulting mass profiles are shown in the fourth row, with the same color coding (blue assuming a constant $M_*/L$, red for radially varying $M_*/L$). Effective radii obtained by integrating the surface density profiles are shown for both profiles, with the hatched areas indicating the $1-\sigma$ errors. The difference between assuming a radially constant mass-to-light ratio and actually accounting for $M_*/L$ variations is clear. Radial $M_*/L$ variations can lead to half-mass radii that are more than a factor 5 smaller than rest-frame $g$ band half-light radii, with errors ranging from more than a factor 2 (first column) to smaller than 10\% (second through fourth columns). These size differences will be discussed in more detail in the next Section. The surface density profiles of all galaxies in our sample are shown in Figure~\ref{fig:profiles}, and are also given in Table~\ref{tab:profiles}. Parameters derived from these profiles are given in Table~\ref{tab:pars}.

\begin{sidewaystable}
\centering
\caption{Galaxy parameters.\label{tab:pars}}
\begin{tabular}{c c c c c c c c c c c c c c}
\tabletypesize{\scriptsize}
\\
\hline \hline
  \colhead{ID$^\mathrm{a}$} & \colhead{$z$} & \colhead{$M_*$$^\mathrm{b}$} & \colhead{$\log$ SSFR$^\mathrm{b}$} & \colhead{$m_u$$^\mathrm{c}$} & \colhead{$m_g$$^\mathrm{c}$} & \colhead{$\nabla u-g$$^\mathrm{d}$} & \colhead{$\nabla \log M_*/L_g$$^\mathrm{d}$} & \colhead{$r_u$} & \colhead{$r_g$} & \colhead{$r_{mass}$} & \colhead{$n_u$} & \colhead{$n_g$} & \colhead{$n_{mass}$} \\
  \colhead{} & \colhead{} & \colhead{($M_\odot$)} & \colhead{($\log$ yr$^{-1}$)} & \colhead{(AB mag)} & \colhead{(AB mag)} & \colhead{} & \colhead{} & \colhead{(kpc)} & \colhead{(kpc)} & \colhead{(kpc} & \colhead{} & \colhead{} & \colhead{} \\
\hline
 981 & 1.13 & 10.97 & -10.22 & $23.65 \pm 0.10$ & $22.01 \pm 0.08$ & $-0.41 \pm 0.06$ & $-0.37 \pm 0.05$ & $ 3.17 \pm 0.22$ & $2.94 \pm 0.09$ & $2.32 \pm 0.10$ & $1.56 \pm 0.04$ & $2.15 \pm 0.10$ & $2.89 \pm 0.29$ \\
1030 & 1.10 & 11.18 & -10.85 & $22.92 \pm 0.10$ & $21.32 \pm 0.08$ & $-0.03 \pm 0.04$ & $-0.03 \pm 0.04$ & $ 2.22 \pm 0.34$ & $1.86 \pm 0.20$ & $1.52 \pm 0.15$ & $4.43 \pm 0.21$ & $5.60 \pm 0.54$ & $7.39 \pm 3.59$ \\
1043 & 1.14 & 11.02 & -10.99 & $23.49 \pm 0.10$ & $21.83 \pm 0.08$ & $-0.79 \pm 0.03$ & $-0.71 \pm 0.03$ & $ 3.10 \pm 0.26$ & $2.20 \pm 0.13$ & $0.79 \pm 0.04$ & $1.96 \pm 0.11$ & $3.10 \pm 0.22$ & $5.13 \pm 1.70$ \\
1060 & 2.35 & 11.17 & -10.16 & $23.84 \pm 0.10$ & $22.27 \pm 0.11$ & $-0.59 \pm 0.05$ & $-0.59 \pm 0.05$ & $ 3.11 \pm 0.17$ & $3.05 \pm 0.16$ & $0.63 \pm 0.12$ & $2.64 \pm 0.59$ & $4.40 \pm 0.15$ & $5.00 \pm 0.01$ \\
1088 & 1.72 & 10.75 & -10.12 & $23.58 \pm 0.08$ & $22.50 \pm 0.09$ & $ 0.18 \pm 0.04$ & $ 0.18 \pm 0.04$ & $ 0.94 \pm 0.06$ & $0.91 \pm 0.03$ & $0.86 \pm 0.10$ & $3.09 \pm 0.23$ & $4.71 \pm 0.63$ & $7.82 \pm 4.12$ \\
1148 & 1.10 & 10.99 & -10.53 & $22.99 \pm 0.10$ & $21.50 \pm 0.08$ & $-0.19 \pm 0.02$ & $-0.17 \pm 0.02$ & $ 2.07 \pm 0.16$ & $1.83 \pm 0.08$ & $1.36 \pm 0.05$ & $2.01 \pm 0.10$ & $2.50 \pm 0.29$ & $3.55 \pm 0.33$ \\
1175 & 1.09 & 10.91 & -10.15 & $23.45 \pm 0.10$ & $21.97 \pm 0.08$ & $-0.29 \pm 0.03$ & $-0.26 \pm 0.03$ & $ 3.34 \pm 0.81$ & $2.40 \pm 0.20$ & $1.78 \pm 0.16$ & $3.39 \pm 0.17$ & $3.44 \pm 0.14$ & $2.92 \pm 0.47$ \\
1190 & 1.10 & 11.10 &  -9.39 & $22.13 \pm 0.10$ & $20.91 \pm 0.08$ & $-0.22 \pm 0.04$ & $-0.20 \pm 0.04$ & $ 5.07 \pm 0.28$ & $4.96 \pm 0.18$ & $4.75 \pm 0.20$ & $1.33 \pm 0.09$ & $1.52 \pm 0.10$ & $1.83 \pm 0.02$ \\
1242 & 1.10 & 11.31 & -10.74 & $22.48 \pm 0.10$ & $21.07 \pm 0.08$ & $-0.15 \pm 0.05$ & $-0.14 \pm 0.04$ & $14.56 \pm 4.58$ & $8.32 \pm 1.57$ & $8.06 \pm 0.94$ & $5.38 \pm 0.25$ & $6.86 \pm 0.54$ & $5.15 \pm 0.87$ \\
1246 & 1.10 & 10.95 &  -9.23 & $22.88 \pm 0.10$ & $21.53 \pm 0.08$ & $-0.24 \pm 0.05$ & $-0.21 \pm 0.05$ & $ 2.79 \pm 0.07$ & $2.67 \pm 0.06$ & $2.23 \pm 0.05$ & $0.94 \pm 0.02$ & $1.32 \pm 0.09$ & $2.64 \pm 0.21$ \\
\nodata & \nodata & \nodata & \nodata & \nodata & \nodata & \nodata & \nodata & \nodata & \nodata & \nodata & \nodata & \nodata & \nodata \\
\hline \\
\multicolumn{14}{l}{This Table is published in its entirety in the electronic edition of this Paper, and can also be downloaded from \url{http://www.strw.leidenuniv.nl/~szomoru/}. A portion is shown here} \\
\multicolumn{14}{l}{~~~for guidance regarding its form and content.} \\
\multicolumn{14}{l}{$^\mathrm{a}$FIREWORKS ID \citep{wuy08} for $0.5 < z < 2.5$ sources, NYU-VAGC ID \citep{bla05} for $z < 0.5$ sources.} \\
\multicolumn{14}{l}{$^\mathrm{b}$Masses and SSFRs are taken from the FIREWORKS catalog for $0.5 < z < 2.5$ sources, and from the MPA-JHU catalogs for $z < 0.5$ sources (see Section~\ref{sec:data} for details).} \\
\multicolumn{14}{l}{$^\mathrm{c}$Total apparent magnitudes in rest-frame filters.} \\
\multicolumn{14}{l}{$^\mathrm{d}$Color and $M_*/L_g$ gradients are defined as $\Delta(u-g)/\Delta(\log r)$ and $\Delta(\log M_*/L_g)/\Delta(\log r)$, respectively. They are calculated using a linear fit to every profile between the PSF HWHM and the} \\
\multicolumn{14}{l}{~~~radius at which errors start to dominate (typically 8-10 kpc).}
\end{tabular}
\end{sidewaystable}

\begin{table*}
\centering
\caption{Stellar mass surface density profiles.\label{tab:profiles}}
\begin{tabular}{c c c c c c c c}
\\
\hline \hline
  \colhead{ID$^\mathrm{a}$} & \colhead{redshift} & \colhead{$r_\mathrm{arcsec}$} & \colhead{$r_\mathrm{kpc}$} & \colhead{$\mu_u$} & \colhead{$\mu_g$} & \colhead{$\log M_*/L_g$} & \colhead{$\log \Sigma$} \\
  \colhead{} & \colhead{} & \colhead{(arcsec)} & \colhead{(kpc)} & \colhead{(AB mag arcsec$^{-2}$)} & \colhead{(AB mag arcsec$^{-2}$)} & \colhead{($\log M_\odot/L_\odot$)} & \colhead{($\log M_\odot$ kpc$^{-2}$)} \\
\hline
981 & 1.13 & 0.0180 & 0.148 & $21.470 \pm 0.0056$ & $19.442 \pm 0.0008$ & $0.671 \pm 0.1301$ & $10.552 \pm 0.1301$ \\
981 & 1.13 & 0.0198 & 0.163 & $21.509 \pm 0.0058$ & $19.499 \pm 0.0009$ & $0.655 \pm 0.1301$ & $10.514 \pm 0.1301$ \\
981 & 1.13 & 0.0216 & 0.177 & $21.547 \pm 0.0063$ & $19.552 \pm 0.0009$ & $0.642 \pm 0.1301$ & $10.479 \pm 0.1301$ \\
981 & 1.13 & 0.0240 & 0.197 & $21.596 \pm 0.0083$ & $19.620 \pm 0.0011$ & $0.625 \pm 0.1302$ & $10.435 \pm 0.1302$ \\
981 & 1.13 & 0.0264 & 0.217 & $21.642 \pm 0.0087$ & $19.683 \pm 0.0015$ & $0.610 \pm 0.1302$ & $10.394 \pm 0.1302$ \\
981 & 1.13 & 0.0288 & 0.237 & $21.687 \pm 0.0090$ & $19.744 \pm 0.0016$ & $0.596 \pm 0.1303$ & $10.356 \pm 0.1303$ \\
981 & 1.13 & 0.0318 & 0.261 & $21.740 \pm 0.0099$ & $19.815 \pm 0.0018$ & $0.580 \pm 0.1303$ & $10.312 \pm 0.1303$ \\
981 & 1.13 & 0.0348 & 0.286 & $21.792 \pm 0.0124$ & $19.883 \pm 0.0019$ & $0.566 \pm 0.1305$ & $10.271 \pm 0.1305$ \\
981 & 1.13 & 0.0384 & 0.315 & $21.851 \pm 0.0131$ & $19.960 \pm 0.0020$ & $0.550 \pm 0.1305$ & $10.224 \pm 0.1305$ \\
981 & 1.13 & 0.0426 & 0.350 & $21.919 \pm 0.0139$ & $20.044 \pm 0.0023$ & $0.535 \pm 0.1306$ & $10.176 \pm 0.1306$ \\
\nodata & \nodata & \nodata & \nodata & \nodata & \nodata & \nodata & \nodata \\
\hline \\
\multicolumn{8}{l}{This Table is published in its entirety in the electronic edition of this Paper, and can also be downloaded from} \\
\multicolumn{8}{l}{~~~\url{http://www.strw.leidenuniv.nl/~szomoru/}. A portion is shown here for guidance regarding its form and content.} \\
\multicolumn{8}{l}{$^\mathrm{a}$FIREWORKS ID \citep{wuy08} for $0.5 < z < 2.5$ sources, NYU-VAGC ID \citep{bla05} for $z < 0.5$ sources.}
\end{tabular}
\end{table*}

\section{Mass-weighted sizes}\label{sec:radii}

\subsection{Structural parameter derivation}

\begin{figure*}
\epsscale{1.2}
\plotone{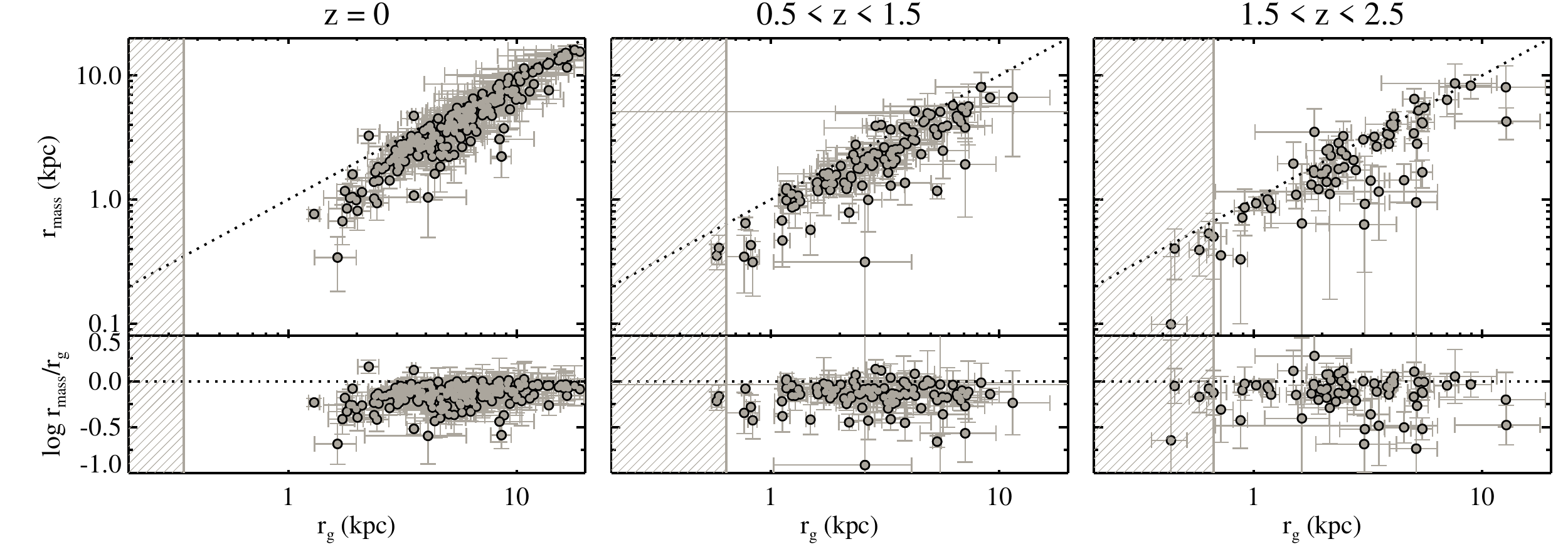}
\caption{Half-mass radii versus rest-frame $g$ band half-light radii, for galaxies with $M_* > 10^{10.7} M_\odot$. The approximate PSF HWHM in each redshift bin is indicated by the hatched region. At all redshifts the half-mass radii of galaxies are on average $25\%$ smaller than their half-light radii. Values as low as $\log{r_{mass}/r_g} = -0.7$ are found.}\label{fig:masslumsize}
\end{figure*}

Having constructed stellar mass surface density profiles, we focus our attention on galaxy sizes. Half-mass radii are amongst the most basic galaxy characteristics and are especially important in the context of galaxy evolution, given their close connection to the build-up of stellar mass. Half-light radii, used in all studies of galaxy size evolution, can be strongly influenced by stellar population differences within a given galaxy. The surface density profiles described in the previous Section provide information on the true mass distribution of these galaxies.

Half-mass radii are calculated by integrating the surface density profiles out to a radius of $\sim100$ kpc. This corresponds to $\sim12$ arcsec for galaxies at $z\sim1 - 2$, and $\sim90$ arcsec for the $z=0$ galaxies. The errors in the half-mass radii (and other parameters derived from the surface density profiles) are estimated in two ways. Firstly, we estimate the uncertainty due to errors in the flux profile by varying the $u-g$ color profiles within their $1-\sigma$ error ranges, deriving mass profiles based on these new color profiles, and calculating the resulting range of half-mass radii. Secondly, we estimate the effects of PSF variations on our size determinations. This is done by rerunning the entire procedure described in Section~\ref{sec:analysis} using twelve different stellar PSFs. The median error in $r_{mass}$ due to these sources is 18\%.

In Figure~\ref{fig:masslumsize} the half-mass radii are plotted against half-light radii measured from the rest-frame $g$ band surface brightness profiles. The half-mass radii are generally smaller than the half-light radii: $\langle \log{r_{mass}/r_g} \rangle \approx -0.12\pm0.01$. Overall, the stellar mass distributions of massive galaxies are more concentrated than their rest-frame optical light distributions by about 25\%. The overall trend is in agreement with previous studies, which have found that galaxies at these redshifts tend to show negative color gradients, such that their cores are relatively red (e.g., \citealt{dok10}; \citealt{szo11}; \citealt{guo11}).

The median differences between mass-weighted and luminosity-weighted parameters are given in Table~\ref{tab:diff}. Mass-weighted S\'ersic indices (obtained by fits to the radial surface density profiles) are on average $\sim 5-20 \%$ larger than luminosity-weighted S\'ersic indices. It should be noted that S\'ersic indices are much more difficult to accurately constrain than effective radii, due to their sensitivity to systematic uncertainties; errors on our mass-weighted S\'ersic indices are approximately twice as large as errors on our mass-weighted effective radii.

We also give the difference between total masses calculated by integrating the stellar mass surface density profiles and total masses from integrated photometry. We find that, on average, masses from resolved photometry are higher than those based on integrated light: the median difference is $5-10\%$ at high redshift, and $\sim30\%$ at $z=0$. Our low-redshift result is in agreement with results presented by \cite{zib09}, who derive resolved stellar mass maps of nine nearby galaxies with a range of morphologies. These authors find that mass estimates from integrated photometry may miss up to 40\% of the total stellar mass compared to estimates obtained by summing resolved mass maps, due to dusty regions being underrepresented in the total flux of galaxies.

This has consequences for integrated galaxy colors, which are often used as a proxy for star formation activity. Dusty, red regions with low flux but high mass densities will be underrepresented in colors based on integrated galaxy photometry. We estimate this effect using total rest-frame $u-g$ colors. Mass-weighted colors are calculated as follows:
\begin{equation}
(u-g)_{mass} = \frac{\int{\frac{f_u(r)}{f_g(r)} \Sigma(r)dr}}{\int{\Sigma(r)dr}},
\end{equation}\label{eq:color}
where $f_u(r)$, $f_g(r)$ and $\Sigma(r)$ are the radial $u$ and $g$ band flux density and radial stellar mass surface density profiles, respectively. On average, mass-weighted $u-g$ colors are redder than luminosity-weighted $u-g$ colors by $0.04-0.06$ magnitudes, indicating that the SSFRs implied by luminosity-weighted $u-g$ colors are slightly overestimated compared to those implied by mass-weighted colors.

\begin{table*}
\centering
\caption{Differences between mass-weighted and luminosity-weighted parameters.\label{tab:diff}}
\begin{tabular}{c c c c c c c c c c c c}
\\
\hline \hline
  \colhead{} & \multicolumn{2}{c}{$\log{r_{mass}} - \log{r_g}$} & & \multicolumn{2}{c}{$\log{n_{mass}} - \log{n_g}$} & & \multicolumn{2}{c}{$\log{M_{mass}} - \log{M_g}^\mathrm{a}$} & & \multicolumn{2}{c}{$(u-g)_{mass} - (u-g)_g^\mathrm{b}$} \\
  \cline{2-3} \cline{5-6} \cline{8-9} \cline{11-12} 
  \colhead{} & \colhead{median} & \colhead{$\sigma_\mathrm{NMAD}$} & & \colhead{median} & \colhead{$\sigma_\mathrm{NMAD}$} & & \colhead{median} & \colhead{$\sigma_\mathrm{NMAD}$} & & \colhead{median} & \colhead{$\sigma_\mathrm{NMAD}$} \\
\hline
$z = 0$ & $-0.12 \pm 0.01$ & 0.08 & & $0.06 \pm 0.03$ & 0.31 & & $0.12 \pm 0.01$ & 0.14 & & $0.059 \pm 0.004$ & 0.048 \\
$0.5 < z < 1.5$ & $-0.14 \pm 0.01$ & 0.11 & & $0.09 \pm 0.02$ & 0.20 & & $0.04 \pm 0.01$ & 0.06 & & $0.044 \pm 0.006$ & 0.048 \\
$1.5 < z < 2.5$ & $-0.10 \pm 0.02$ & 0.13 & & $0.02 \pm 0.03$ & 0.19 & & $0.02 \pm 0.01$ & 0.07 & & $0.043 \pm 0.007$ & 0.051 \\
\hline \\
\multicolumn{12}{l}{Note: the $mass$ subscript indicates parameters derived using the true mass profiles (i.e., with radially varying $M_*/L$), while the $g$ subscript} \\
\multicolumn{12}{l}{~~~indicates parameters derived using profiles that assume a constant $M_*/L$ profile (i.e., equivalent to the rest-frame $g$ band profiles).} \\
\multicolumn{12}{l}{$^\mathrm{a}$$\log{M_{mass}}$ is the total stellar mass derived by summing the resolved mass density information. $\log{M_g}$ is the total stellar mass based on} \\
\multicolumn{12}{l}{~~~integrated photometry.} \\
\multicolumn{12}{l}{$^\mathrm{b}$$(u-g)_{mass}$ is the mass-weighted color (see Equation~\ref{eq:color}), and $(u-g)_g$ is the luminosity-weighted color.}
\end{tabular}
\end{table*}

\subsection{Sources of uncertainty}

The stellar mass surface density profiles are generally not very sensitive to the extrapolation of the color profiles towards larger radii, due to the fact that the total flux at $r > 10$ kpc is very low. Furthermore, this source of uncertainty is explicitly included in our errors, since they are estimated by varying the $u-g$ color that is used for the extrapolation. Uncertainty in the shape of the color profile within the PSF HWHM is more difficult to quantify, and may be especially important for the smallest galaxies ($r_g < 1$ kpc). In order to estimate the stability of our derived sizes we have explored several alternative approaches: one in which the inner $u-g$ color is kept fixed to the integrated value within 1 kpc; and one in which the best-fit linear color gradient is extrapolated from the PSF HWHM inward. Neither approach alters our results.

A possible concern is regions with very high dust content, which could potentially obscure stellar light to such a degree that our $u-g$ --- $M_*/L$ conversion becomes ineffective, simply because all stellar light is extincted. Very high central dust concentrations may result in an underestimate of the inner mass content of galaxies, and an overestimate of their half-mass radii. Such effects are a significant source of systematic uncertainty in our analysis, and can only be quantified by measuring the light reemitted by dust. This requires infrared imaging at \textit{HST} resolution or better, which will become possible in the near future using ALMA.

\begin{figure}
\epsscale{1.2}
\plotone{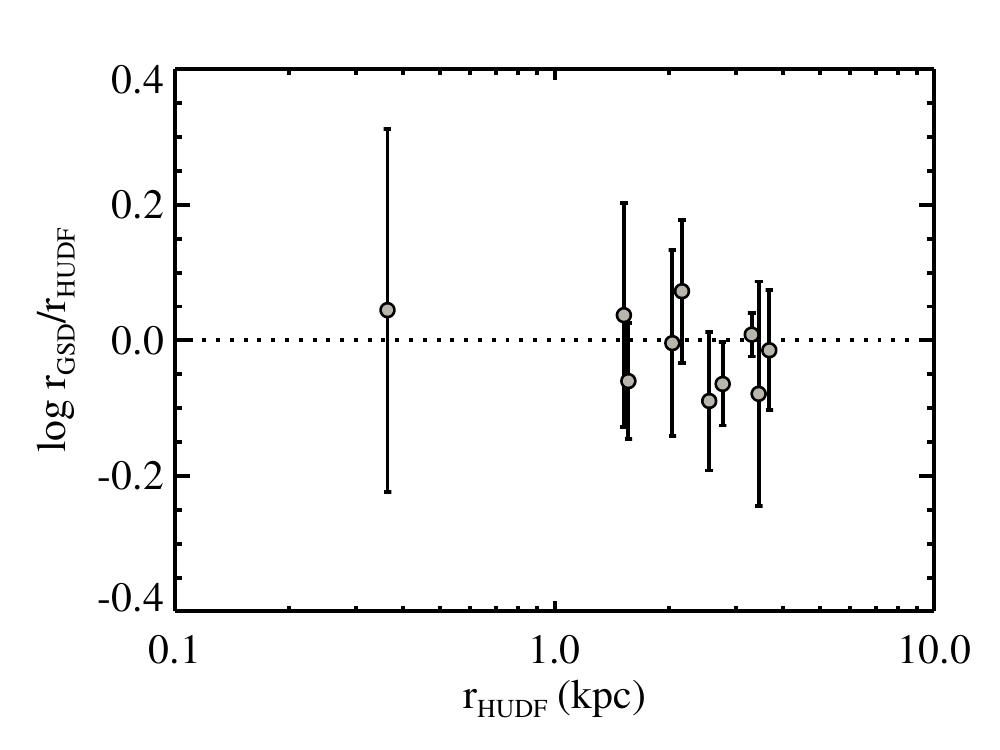}
\caption{Comparison of half-mass radii derived using CANDELS GOODS-South data and HUDF data, for the galaxies with $0.5 < z < 2.5$ and $M_* > 10^{10.7}$ that are found in both datasets. The two datasets agree within the errors.}\label{fig:hudf}
\end{figure}

The measured shape and radial extent of a galaxy's surface brightness profile are sensitive to the depth to which the galaxy is imaged. A lack of imaging depth can result in errors in sky background estimation, as well as a portion of the galaxy's emission being lost in the background noise. Generally, low-S/N data is likely to introduce systematic effects, such that measured sizes and S\'ersic indices are smaller than the true values (e.g., \citealt{tru06}; \citealt{wil10}). As a consistency check we therefore compare our results to those based on ultradeep optical and NIR data acquired over the HUDF (\citealt{bec06}; \citealt{bou11}). These data are $\sim2$ magnitudes deeper in the NIR than the CANDELS data and should therefore be unaffected by these surface brightness effects. We find 11 galaxies with $M_* > 10^{10.7} M_\odot$ and $0.5 < z < 2.5$ which are imaged by both CANDELS and HUDF. The HUDF WFC3 data of one of these galaxies shows some artificial background features, which are the result of the background subtraction process; this galaxy is therefore excluded from the comparison. The HUDF imaging is processed in exactly the same way as described in Section~\ref{sec:analysis} for the CANDELS imaging. The resulting half-mass radii are compared to the CANDELS half-mass radii in Figure~\ref{fig:hudf}. The correspondence is good; HUDF and CANDELS measurements of $r_{mass}$ lie within $1 \sigma$ of each other, and we find no systematic offset. Thus, based on this subsample of galaxies we conclude that the CANDELS data are sufficiently deep for our purposes.

\subsection{Evolution with redshift}

\begin{figure*}
\epsscale{1.2}
\plotone{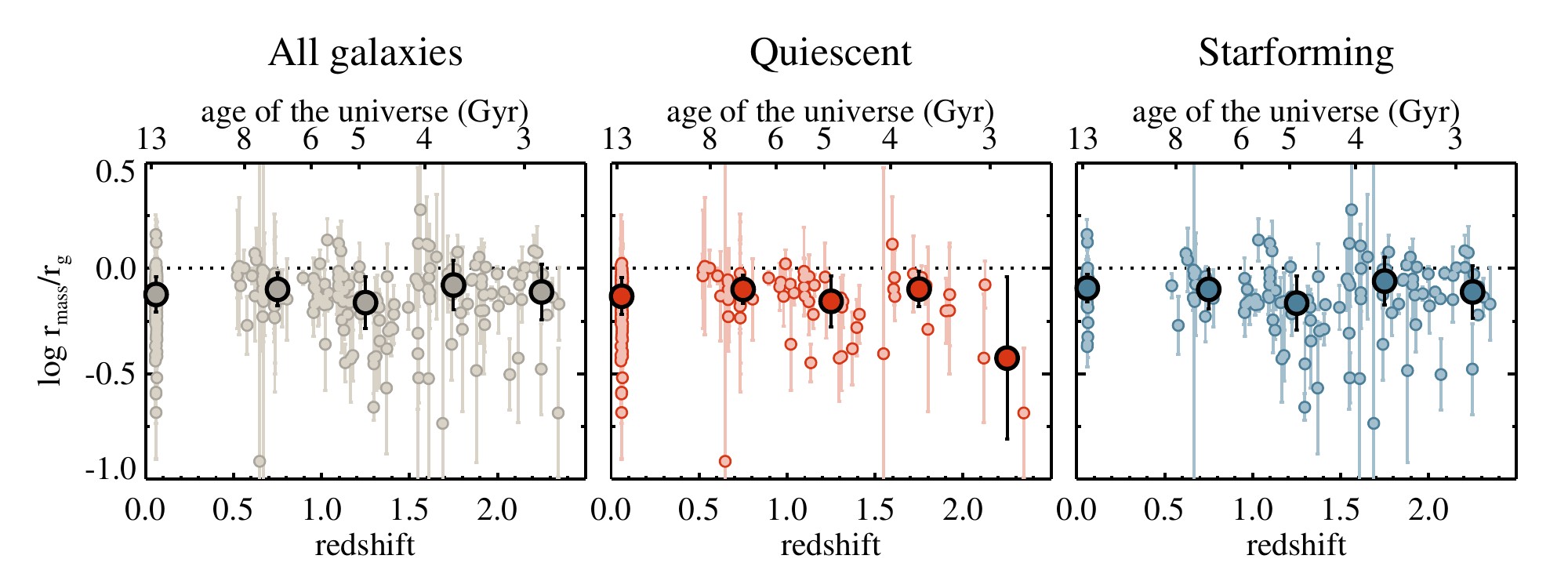}
\caption{The ratio of half-mass size to half-light size, $r_{mass}/r_g$, as a function of redshift, for galaxies with $M_* > 10^{10.7}$. From left to right are plotted all galaxies, quiescent galaxies, and starforming galaxies. The large symbols indicate the running median, with the error bars indicating the $1-\sigma$ scatter of the distribution. Our results indicate that, at fixed mass, galaxies have similar $r_{mass}/r_g$ values at all redshifts between $z=0$ and $z=2.5$.}\label{fig:sizediff}
\end{figure*}

Figure~\ref{fig:masslumsize} indicates that mass distributions are on average more concentrated than light distributions at all redshifts. We now investigate the redshift evolution of this concentration difference in more detail. In Figure~\ref{fig:sizediff} we plot $r_{mass}/r_g$ as a function of redshift for all galaxies in our sample, as well as for quiescent and starforming galaxies separately. Quiescent galaxies are defined to have SSFR $< 0.3/t_H$, where $t_H$ is the Hubble time. Individual galaxies are indicated by small circles, and median values for each redshift bin are shown as large, darker circles. The error bars on the median values indicate the scatter of the distribution.

Overall, the difference between mass-weighted and luminosity-weighted radius does not seem to evolve with redshift: $\log{r_{mass}/r_g} = -0.12\pm0.01$ at $z=0$, $-0.14\pm0.01$ at $z\sim1$, and $-0.10\pm0.02$ at $z\sim2$ (see also Table~\ref{tab:table}). Similarly, the values for starforming galaxies are consistent with no evolution with redshift. For quiescent galaxies, there is a hint of decreasing values of $\log{r_{mass}/r_g}$ at higher redshifts, although the large scatter at $z\sim2$ means that these results are also consistent with zero redshift evolution.

The lack of evolution with redshift agrees with the color gradients presented in \cite{szo11}. This study showed that the radial color gradients in galaxies with $10^{10} < M_*/M_\odot < 10^{11}$ are nearly constant between $0 < z < 2.5$. This seems counterintuitive, since one would expect bulge growth in galaxies to result in steeper color gradients at lower redshift. However, it should be stressed that the comparison in Figure~\ref{fig:sizediff} is between galaxies of the same mass. It is therefore not a comparison between high-redshift galaxies and their descendants, but rather a comparison between high-redshift galaxies and their low-redshift analogs. Thus, $z\sim2$ galaxies have similar color gradients as low redshift galaxies of the same mass. This holds for starforming and quiescent galaxies seperately, and for the entire population as a whole.

\subsection{Dependence on galaxy parameters}

\begin{figure*}
\epsscale{1.2}
\plotone{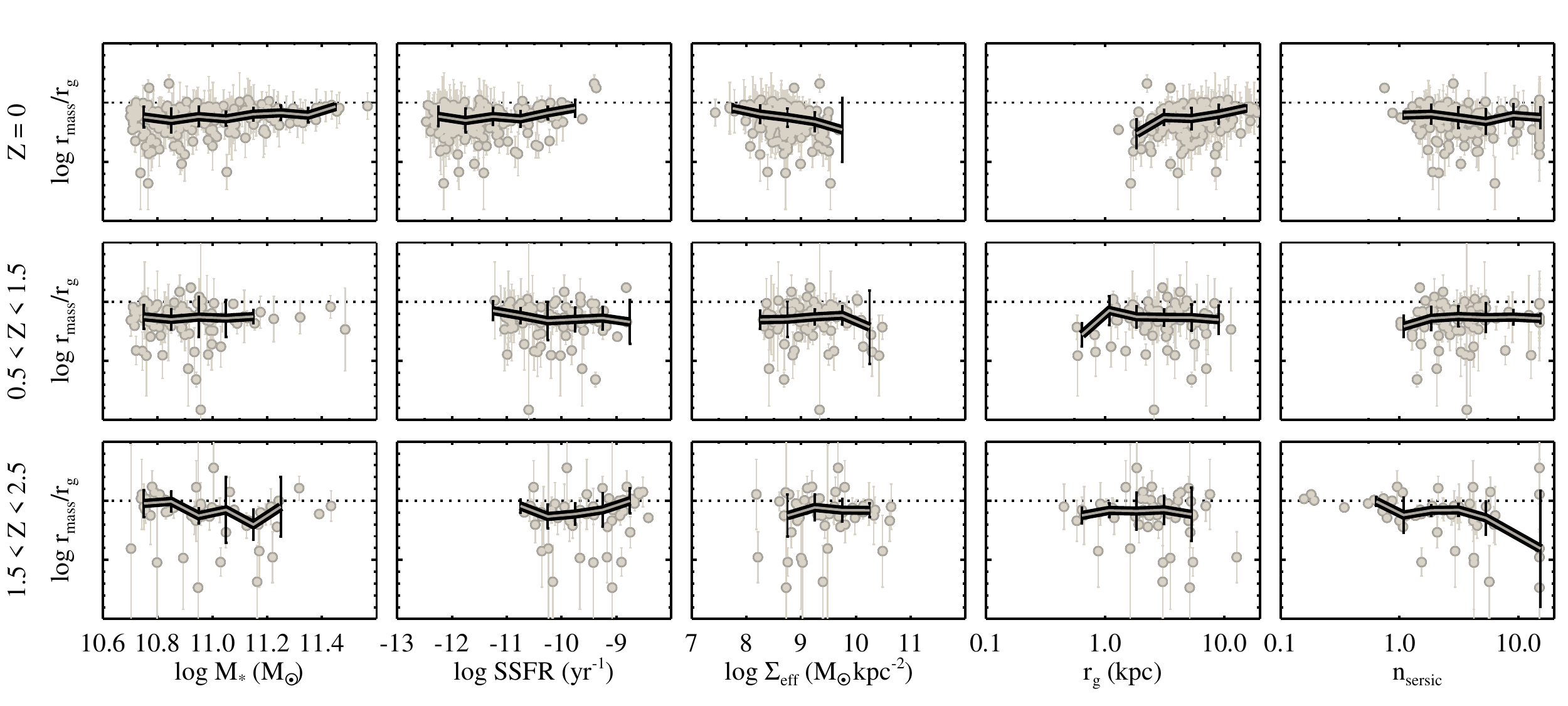}
\caption{The correlations between $r_{mass}/r_g$ and, from left to right: total stellar mass, specific star formation rate, and structural parameters measured in the rest-frame $g$ band (effective surface density, effective radius, and S\'ersic index). Individual galaxy measurements are indicated by circles, and running medians are indicated by the thick lines. The error bars on the running medians indicate the $1-\sigma$ scatter. Each row corresponds to a redshift bin. $r_{mass}/r_g$ correlates weakly, if at all, with starforming activity at both high and low redshift.}\label{fig:corr}
\end{figure*}

The transition of galaxies from the starforming to the passive population is coupled to changes in almost all aspects of their structure and morphology (e.g., \citealt{kau03};\citealt{tof07}; \citealt{fra08}; \citealt{bel08}; \citealt{dok11}; \citealt{szo11}; \citealt{wuy11}; \citealt{bel12}). The growth of a bulge, in particular, implies that color gradients should steepen as galaxies are quenched. We therefore expect some degree of correlation between $r_{mass}/r_g$ and the galaxy parameters that correlate with star forming activity, such as SSFR, size, S\'ersic index, and effective surface density. We investigate these correlations in Figure~\ref{fig:corr}. Individual galaxies are indicated by small light-colored circles, and the running medians in each panel are indicated by darker lines.

Somewhat surprisingly, the median value of $r_{mass}/r_g$ is close to constant as a function of galaxy parameters in all redshift bins. There is some evidence for a trend between $r_{mass}/r_g$ and galaxy morphology, such that $r_{mass}/r_g$ is smaller for galaxies with low SSFR, small sizes and high S\'ersic indices. However, this trend is very weak. Overall, the correlation coefficients at $z\sim1$ and $z\sim2$ are very low and not significant ($r \sim 0.01$, $p \sim 0.4$). The exceptions to this are the correlations between $\log{r_{mass}/r_g}$ and $\log{\mathrm{SSFR}}$ and $n_{sersic}$, at $z\sim2$; although these are also quite low, they are significant ($r \sim 0.35$, $p < 0.01$). Similarly, the correlation coefficients at $z=0$ are low, but significant ($r \sim 0.2 - 0.3$, $p < 0.01$). We conclude that the difference between half-mass size and half-light size correlates very weakly with galaxy structure and star forming activity.

\begin{figure*}
\centering
\epsscale{1.1}
\plotone{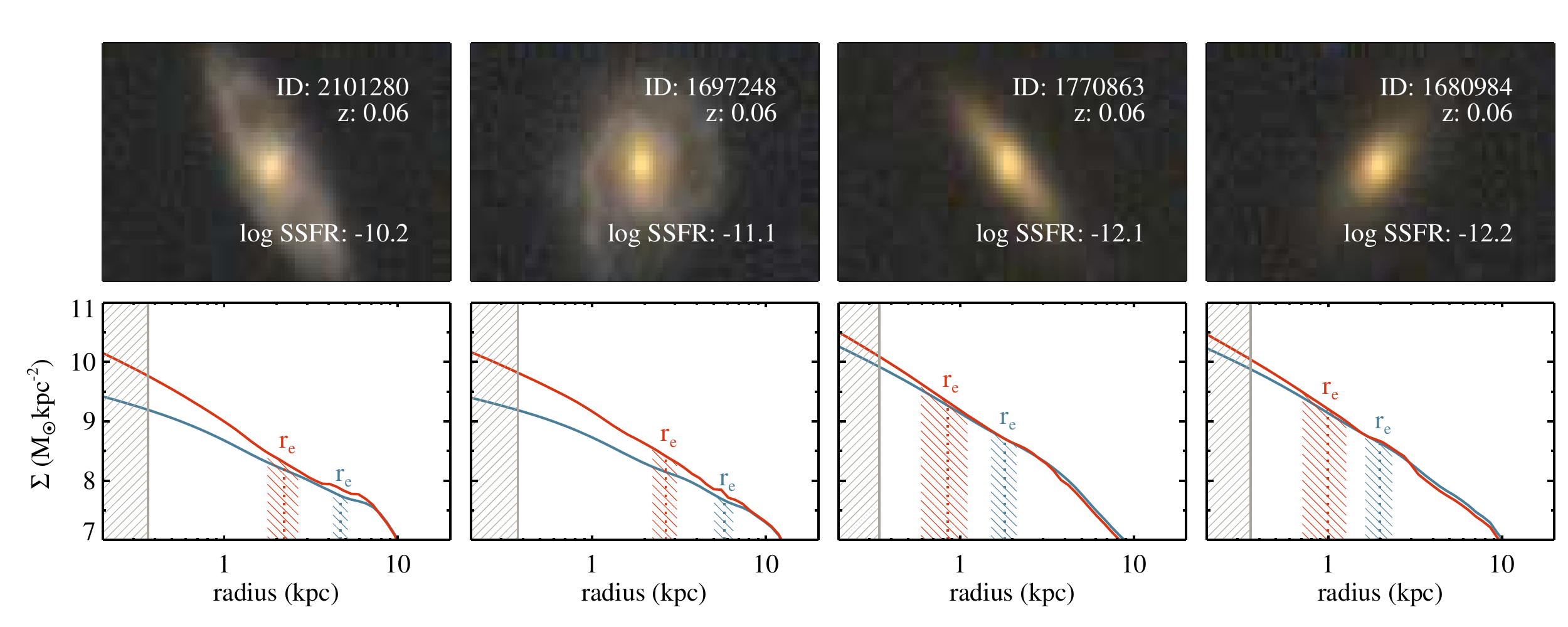}
\plotone{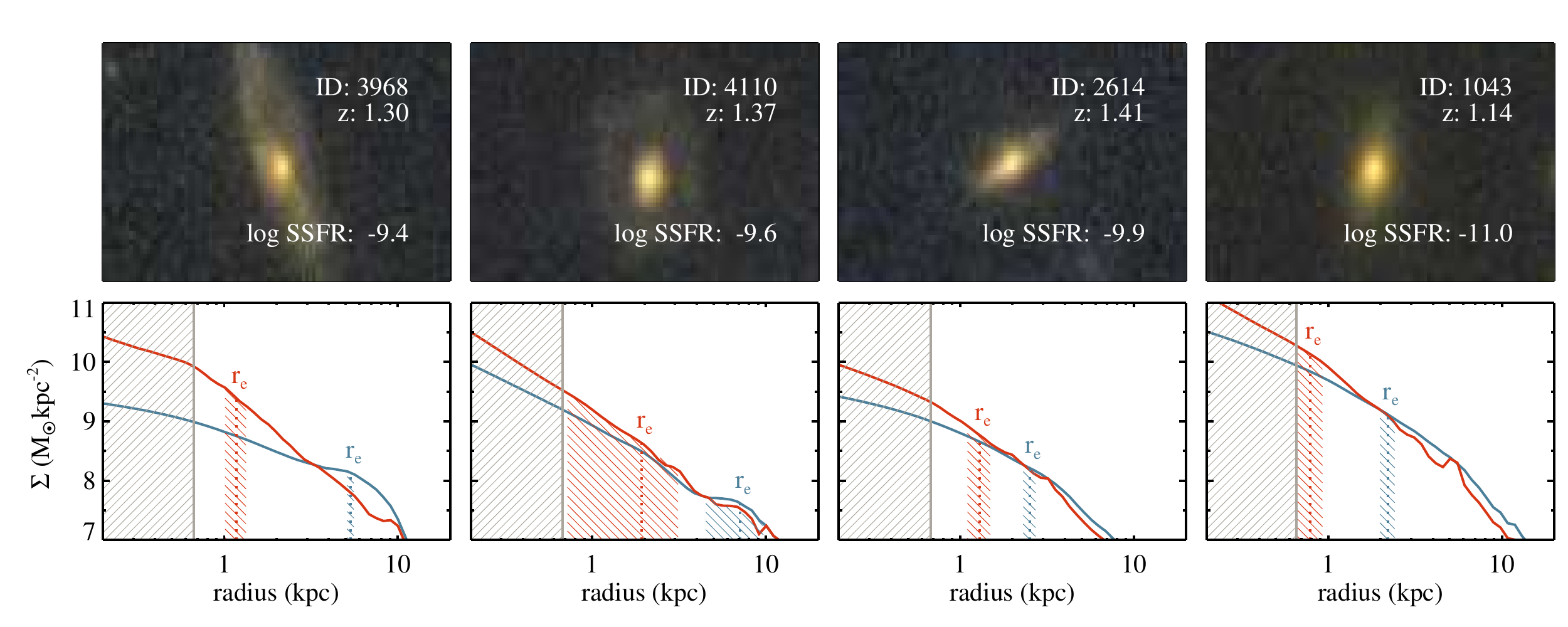}
\plotone{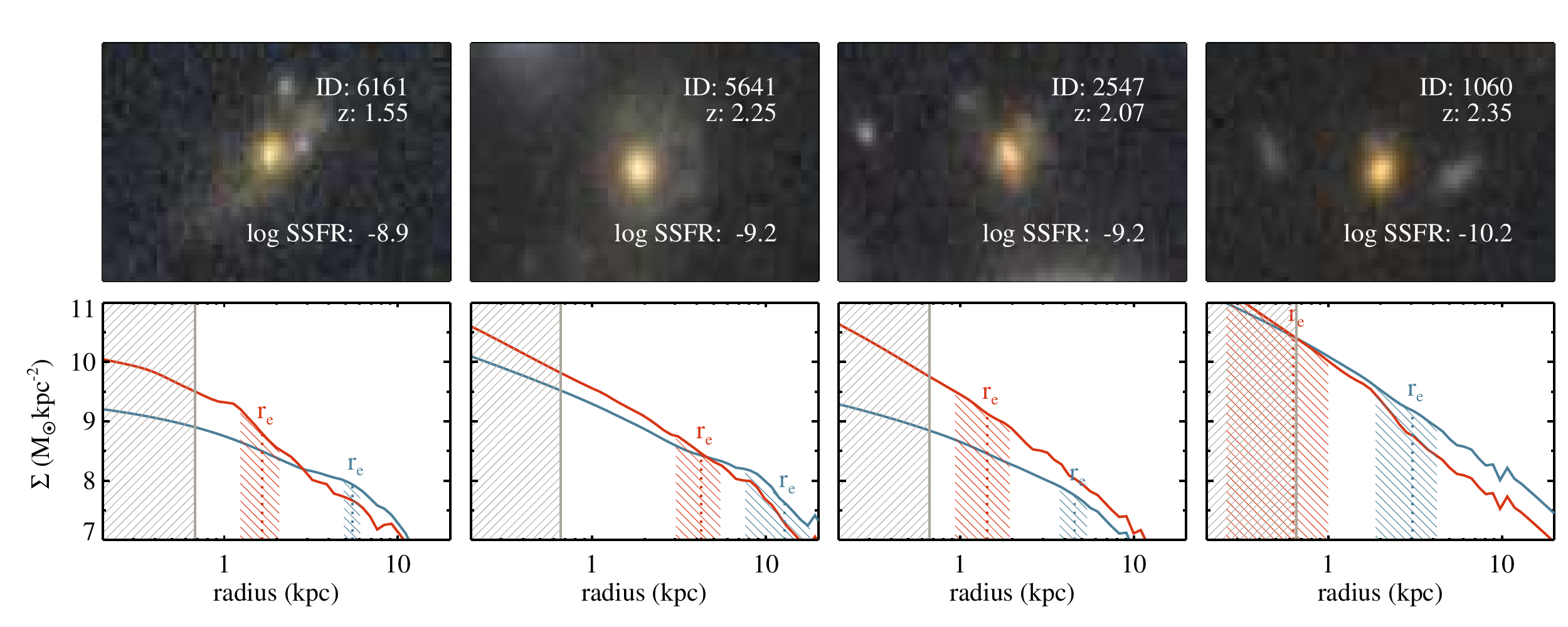}
\caption{A representative selection of galaxies with $r_{mass}/r_g < 0.5$, ordered by SSFR. All galaxies are extended, disk-like galaxies with large central bulges. Measurements in the rest-frame $g$ band result in severely overestimated sizes compared to the half-mass sizes. The color gradients are measured with high precision; the large differences between $r_{mass}$ and $r_g$ are therefore not the result of measurement errors.}\label{fig:extreme}
\end{figure*}

Although the average difference between half-mass and half-light radius is remarkably constant, for some galaxies the difference is up to a factor eight. These large differences warrant closer inspection. In Figure~\ref{fig:extreme} we show images and profiles of a number of galaxies with $r_{mass}/r_g < 0.5$. Four representative galaxies have been selected from each redshift bin, and are plotted in order of decreasing SSFR. The flux profile errors of the galaxies shown are typical for the galaxies in this sample. Nearly all of these galaxies are dominated by large central bulges, but also contain extended, blue disks. These large galaxies have very well-measured surface brightness profiles; errors are low, and the profiles are measured accurately out to very large radii. We are therefore confident that the color gradients, although very steep, are real.

This group of galaxies demonstrates the importance of obtaining resolved stellar mass information. Measurements made at rest-frame optical wavelengths severely underestimate the relative importance of the central bulge component in these galaxies, due to the high luminosity of the blue disks surrounding them. As a result, for a sizeable fraction ($\sim10\%$) of massive galaxies the structure as measured at rest-frame optical wavelengths severely misrepresents the mass distribution in these galaxies.

This population of large, bulge-dominated disk galaxies is very interesting. They have stellar masses up to $\sim10^{11} M_\odot$ and are distributed across the entire range of measured SSFRs. Moving from high to low SSFR, we can see a progression from Sa or Sb-like systems to objects that look more like S0s and elliptical galaxies. This is most evident at low redshift, where galaxies have large angular sizes. The color difference between the central bulge and extended disk gradually decreases from high- to low-SSFR galaxies. In other words, it seems that we are observing these galaxies during the phase where a central bulge has recently formed and star formation is turning off. It is unclear why other galaxies with similar SSFRs do not contain such large bulges; this may depend on the dark matter halos in which they are located, or may perhaps be due to different quenching processes.

\section{Summary and conclusions}

We have presented the first consistently measured stellar mass surface density profiles for individual galaxies at redshifts between $z=0$ and $z=2.5$. These profiles have been derived using an empirical relation between rest-frame color and stellar mass-to-light ratio. This simple method does not yield the same detailed information regarding, e.g., stellar ages and dust content as resolved SED-fitting techniques. However, it is robust to variations in stellar population properties; changes in stellar age, metallicity, or dust content shift galaxies roughly along the empirical relation, and are thus implicitly included in our $M_*/L$ determination.

The key advantage of this study compared to previous work at high redshift (e.g., \citealt{wuy12}; \citealt{lan12}) is the fact that the profiles presented here are deconvolved for PSF smoothing. This is crucial for measurements of high-redshift galaxy structure, since these distant and often physically small galaxies have angular sizes that are in many cases comparable to the \textit{HST} PSF size (e.g., \citealt{dad05}; \citealt{tru06}; \citealt{tof07}; \citealt{dok08}; \citealt{szo10}; \citealt{cas10}; \citealt{szo12}). Since our surface density profiles are derived from deconvolved surface brightness profiles, they can be used to correctly measure structural parameters such as sizes and S\'ersic indices.

The considerable depth of the data used in this study allows us to probe galaxy fluxes and colors out to large radii. The robustness of the resulting structural parameters has been tested using ultradeep data taken over the HUDF, which overlaps with the CANDELS GOODS-South field. A comparison of galaxy parameters derived using the two datasets confirms that our results are not systematically affected by surface brightness effects.

We have shown that the half-mass radii of galaxies between $z=0$ and $z=2.5$ are on average 25\% smaller than their rest-frame optical half-light radii. There is significant scatter in this size difference, with some galaxies having half-mass radii that are almost an order of magnitude smaller than their half-light radii. We find that, on average, this size difference does not vary with redshift for galaxies at fixed mass. This holds for the population as a whole, as well as for the quiescent and starforming subpopulations seperately. This is an interesting result, as it implies that $z\sim2$ galaxies have similar color gradients as their low-redshift analogs, despite the fact that these low-redshift galaxies formed at a different epoch, and perhaps through very different formation mechanisms.

There does not seem to be a strong correlation between galaxy morphology or star forming activity and the difference between half-mass size and half-light size. However, we do find that the galaxies with the most extreme size differences are almost all extended disk galaxies with very prominent central bulges. These galaxies range from strongly starforming to almost completely quiescent, and may represent a short transitional phase during which the central bulge is prominent and the starforming disk is very young.

There is significant scatter around the empirical relation used to convert $u-g$ color to $M_*/L$, which could lead to a systematic underestimate of the mass-to-light ratios in galaxy regions that are relatively metal-rich or old. However, this effect is likely small for moderate stellar population variations. Similarly, very high central dust concentrations may result in an underestimate of the inner mass content of galaxies, biasing our results towards larger half-mass radii. The high-resolution infrared data needed in order to quantify such dust effects will be available in the (near) future, with instruments such as ALMA.

Inside-out galaxy growth, as described by, e.g., \cite{dok10}, implies that the growth of the most massive galaxies since $z\sim2$ is largely due to material being accreted onto the outer regions of these galaxies. The cores of massive galaxies likely formed in short, violent bursts at higher redshift, and should therefore have star formation histories and stellar populations that are quite different from those in the outer regions. The results presented in this Paper broadly agree with such a picture; the central regions of massive galaxies are redder, and therefore likely older, than the outer regions. Using the method presented in this Paper we cannot, however, disentangle dust, age and metallicity gradients; nor can we constrain the star formation histories within our galaxies. First steps towards a better understanding of stellar population variations within high-redshift galaxies have been made by several authors. Results based on photometry of early-type galaxies (e.g., \citealt{guo11}; \citealt{gar12}) indicate that stellar age and metallicity are the dominant drivers of radial color gradients. Studies based on spectroscopic measurements of gravitationally lensed high-redshift galaxies (e.g., \citealt{cre10}; \citealt{jon10}; \citealt{yua11}; \citealt{que12}; \citealt{jon12}) have shown that most of these galaxies have negative metallicity gradients. These results seem to be roughly consistent with each other, but are based on very differently selected, and rather small, galaxy samples. A broader, more in-depth analysis of radial stellar population variations, for a well-defined sample of starforming galaxies as well as quiescent galaxies, could provide valuable insights into the processes which have shaped the structure of galaxies today.

\acknowledgments

We gratefully acknowledge funding from ERC grant HIGHZ no. 227749. This work is based on observations taken by the CANDELS Multi-Cycle Treasury Program with the NASA/ESA HST, which is operated by the Association of Universities for Research in Astronomy, Inc., under NASA contract NAS5-26555.

This publication makes use of the Sloan Digital Sky Survey (SDSS). Funding for the creation and distribution of the SDSS Archive has been provided by the Alfred P. Sloan Foundation, the Participating Institutions, the National Aeronautics and Space Administration, the National Science Foundation, the U.S. Department of Energy, the Japanese Monbukagakusho, and the Max Planck Society. The SDSS Web site is http://www.sdss.org/. The SDSS Participating Institutions are the University of Chicago, Fermilab, the Institute for Advanced Study, the Japan Participation Group, Johns Hopkins University, the Max Planck Institut fur Astronomie, the Max Planck Institut fur Astrophysik, New Mexico State University, Princeton University, the United States Naval Observatory, and the University of Washington.


\begin{thebibliography}{}
\bibitem[Abazajian et al.(2009)]{aba09} Abazajian, K.~N., Adelman-McCarthy, J.~K., Ag{\"u}eros, M.~A., et al.\ 2009, \apjs, 182, 543
\bibitem[Annis et al.(2011)]{ann11} Annis, J., Soares-Santos, M., Strauss, M.~A., et al.\ 2011, arXiv:1111.6619
\bibitem[Beckwith et al.(2006)]{bec06} Beckwith, S.~V.~W., Stiavelli, M., Koekemoer, A.~M., et al.\ 2006, \aj, 132, 1729
\bibitem[Bell \& de Jong(2001)]{bel01} Bell, E.~F., \& de Jong, R.~S.\ 2001, \apj, 550, 212
\bibitem[Bell(2008)]{bel08} Bell, E.~F.\ 2008, \apj, 682, 355
\bibitem[Bell et al.(2012)]{bel12} Bell, E.~F., van der Wel, A., Papovich, C., et al.\ 2012, \apj, 753, 167
\bibitem[Blanton et al.(2005)]{bla05} Blanton, M.~R., Schlegel, D.~J., Strauss, M.~A., et al.\ 2005, \aj, 129, 2562
\bibitem[Bouwens et al.(2011)]{bou11} Bouwens, R.~J., Illingworth, G.~D., Oesch, P.~A., et al.\ 2011, \apj, 737, 90
\bibitem[Brammer et al.(2008)]{bra08} Brammer, G.~B., van Dokkum, P.~G., \& Coppi, P.\ 2008, \apj, 686, 1503
\bibitem[Brinchmann et al.(2004)]{bri04} Brinchmann, J., Charlot, S., White, S.~D.~M., et al.\ 2004, \mnras, 351, 1151
\bibitem[Bruzual \& Charlot(2003)]{bru03} Bruzual, G., \& Charlot, S.\ 2003, \mnras, 344, 1000
\bibitem[Cameron et al.(2011)]{cam11} Cameron, E., Carollo, C.~M., Oesch, P.~A., et al.\ 2011, \apj, 743, 146
\bibitem[Cassata et al.(2010)]{cas10} Cassata, P., Giavalisco, M., Guo, Y., et al.\ 2010, \apjl, 714, L79
\bibitem[Cresci et al.(2010)]{cre10} Cresci, G., Mannucci, F., Maiolino, R., et al.\ 2010, \nat, 467, 811
\bibitem[Daddi et al.(2005)]{dad05} Daddi, E., Renzini, A., Pirzkal, N., et al.\ 2005, \apj, 626, 680
\bibitem[Franx et al.(2008)]{fra08} Franx, M., van Dokkum, P.~G., Schreiber, N.~M.~F., et al.\ 2008, \apj, 688, 770
\bibitem[Gargiulo et al.(2012)]{gar12} Gargiulo, A., Saracco, P., Longhetti, M., La Barbera, F., \& Tamburri, S.\ 2012, arXiv:1207.2295
\bibitem[Giavalisco et al.(2004)]{gia04} Giavalisco, M., Ferguson, H.~C., Koekemoer, A.~M., et al.\ 2004, \apjl, 600, L93
\bibitem[Grogin et al.(2011)]{gro11} Grogin, N.~A., Kocevski, D.~D., Faber, S.~M., et al.\ 2011, \apjs, 197, 35
\bibitem[Guo et al.(2011)]{guo11} Guo, Y., Giavalisco, M., Cassata, P., et al.\ 2011, \apj, 735, 18
\bibitem[Jones et al.(2010)]{jon10} Jones, T., Ellis, R., Jullo, E., \& Richard, J.\ 2010, \apjl, 725, L176
\bibitem[Jones et al.(2012)]{jon12} Jones, T., Ellis, R.~S., Richard, J., \& Jullo, E.\ 2012, arXiv:1207.4489
\bibitem[Kauffmann et al.(2003)]{kau03} Kauffmann, G., Heckman, T.~M., White, S.~D.~M., et al.\ 2003, \mnras, 341, 54
\bibitem[Koekemoer et al.(2011)]{koe11} Koekemoer, A.~M., Faber, S.~M., Ferguson, H.~C., et al.\ 2011, \apjs, 197, 36
\bibitem[Labb{\'e} et al.(2003)]{lab03} Labb{\'e}, I., Rudnick, G., Franx, M., et al.\ 2003, \apjl, 591, L95
\bibitem[Lanyon-Foster et al.(2012)]{lan12} Lanyon-Foster, M.~M., Conselice, C.~J., \& Merrifield, M.~R.\ 2012, \mnras, 3340
\bibitem[Peng et al.(2002)]{pen02} Peng, C.~Y., Ho, L.~C., Impey, C.~D., \& Rix, H.-W.\ 2002, \aj, 124, 266
\bibitem[Queyrel et al.(2012)]{que12} Queyrel, J., Contini, T., Kissler-Patig, M., et al.\ 2012, \aap, 539, A93
\bibitem[Szomoru et al.(2010)]{szo10} Szomoru, D., Franx, M., van Dokkum, P.~G., et al.\ 2010, \apjl, 714, L244
\bibitem[Szomoru et al.(2011)]{szo11} Szomoru, D., Franx, M., Bouwens, R.~J., et al.\ 2011, \apjl, 735, L22
\bibitem[Szomoru et al.(2012)]{szo12} Szomoru, D., Franx, M., \& van Dokkum, P.~G.\ 2012, \apj, 749, 121
\bibitem[Taylor et al.(2009)]{tay09} Taylor, E.~N., Franx, M., van Dokkum, P.~G., et al.\ 2009, \apjs, 183, 295
\bibitem[Toft et al.(2005)]{tof05} Toft, S., van Dokkum, P., Franx, M., et al.\ 2005, \apjl, 624, L9
\bibitem[Toft et al.(2007)]{tof07} Toft, S., van Dokkum, P., Franx, M., et al.\ 2007, \apj, 671, 285
\bibitem[Trujillo et al.(2006)]{tru06} Trujillo, I., F{\"o}rster Schreiber, N.~M., Rudnick, G., et al.\ 2006, \apj, 650, 18
\bibitem[van Dokkum et al.(2008)]{dok08} van Dokkum, P.~G., Franx, M., Kriek, M., et al.\ 2008, \apjl, 677, L5
\bibitem[van Dokkum et al.(2010)]{dok10} van Dokkum, P.~G., Whitaker, K.~E., Brammer, G., et al.\ 2010, \apj, 709, 1018
\bibitem[van Dokkum et al.(2011)]{dok11} van Dokkum, P.~G., Brammer, G., Fumagalli, M., et al.\ 2011, \apjl, 743, L15
\bibitem[Williams et al.(2010)]{wil10} Williams, R.~J., Quadri, R.~F., Franx, M., et al.\ 2010, \apj, 713, 738
\bibitem[Wuyts et al.(2008)]{wuy08} Wuyts, S., Labb{\'e}, I., Schreiber, N.~M.~F., et al.\ 2008, \apj, 682, 985
\bibitem[Wuyts et al.(2009)]{wuy09} Wuyts, S., Franx, M., Cox, T.~J., et al.\ 2009, \apj, 700, 799
\bibitem[Wuyts et al.(2011)]{wuy11} Wuyts, S., F{\"o}rster Schreiber, N.~M., van der Wel, A., et al.\ 2011, \apj, 742, 96
\bibitem[Wuyts et al.(2012)]{wuy12} Wuyts, S., F{\"o}rster Schreiber, N.~M., Genzel, R., et al.\ 2012, \apj, 753, 114
\bibitem[Yuan et al.(2011)]{yua11} Yuan, T.-T., Kewley, L.~J., Swinbank, A.~M., Richard, J., \& Livermore, R.~C.\ 2011, \apjl, 732, L14
\bibitem[Zibetti et al.(2009)]{zib09} Zibetti, S., Charlot, S., \& Rix, H.-W.\ 2009, \mnras, 400, 1181
\end{thebibliography}
\end{document}